\documentclass[aps,preprint,showpacs,amssymb]{revtex4}
\usepackage{graphicx}
\usepackage{color}

\begin{document}
\date{\today}
\title{
Heat distribution function for motion in a general potential at low
temperature}

\author{Hans C. Fogedby}
\email{fogedby@phys.au.dk}

\affiliation{Department of Physics and Astronomy, University of
Aarhus\\Ny Munkegade, 8000, Aarhus C, Denmark\\}

\affiliation{Niels Bohr Institute\\
Blegdamsvej 17, 2100, Copenhagen {\O}, Denmark}

\author{Alberto Imparato}
\email{imparato@phys.au.dk}
\affiliation{Department of Physics and
Astronomy, University of Aarhus\\Ny Munkegade, 8000, Aarhus C,
Denmark}
\begin{abstract}
We consider the 1D motion of an overdamped Brownian particle in a
general potential in the low temperature limit. We derive an
explicit expression for the probability distribution for the heat
transferred to the particle. We find that the local minima in the
potential yield divergent side bands in the heat distribution in
addition to the divergent central peak. The position of the bands
are determined by the potential gaps. We, moreover, determine the
tails of the heat distribution.
\end{abstract}
.

\maketitle

There is a strong current interest in the thermodynamics and
statistical mechanics of small fluctuating systems in contact with a
heat reservoir and driven by external forces. The strong interest
stems from the recent possibility of direct manipulation of nano
systems and biomolecules. These techniques permit direct
experimental access to the probability distribution functions (PDFs)
for the work or for the heat exchanged with the environment
\cite{Felix04,Felix05,Seifert06a,Seifert06b,
Wang02,AI07,Ciliberto06,Ciliberto07,Ciliberto08}. These techniques
have also opened the way to the experimental verification of the
fluctuation theorems, which relate the probability of observing
entropy--generating trajectories, with that of observing
entropy-consuming trajectories.
\cite{Jarzynski97,Kurchan98,Gallavotti96,Crooks99,Crooks00,
Seifert05a,Seifert05b,Evans93,Evans94,GC95,LS99,Gaspard04,AI06,
vanZon03,vanZon04,vanZon03a,vanZon04a,Seifert05c}. However, from a
theoretical point of view the time evolution of the work and heat
PDFs for a Brownian particle in an external potential is governed by
a complex partial differential equation \cite{AI07,Seifert05c} whose
explicit solution is only available for simple potentials. It is
therefore of interest to extract some general properties regarding
the motion of a Brownian particle.

In the present letter we consider a Brownian particle in a general
static potential. We show that an explicit asymptotic expression for
the heat PDF can be obtained in the low temperature - long time
limit. As a starting point we consider the stochastic motion in 1D
of an overdamped Brownian particle in the general static potential
$U(x)$. Within a conventional Langevin equation description the
motion is governed by
\begin{eqnarray}
\frac{dx}{dt}=-\frac{\partial U}{\partial x}+\xi,
\label{lan}
\end{eqnarray}
where the noise characterizing the fluctuations imparted by the heat
bath at temperature $T$ is correlated according to
\begin{eqnarray}
\langle\xi(t)\xi(0)\rangle=2T\delta(t). \label{noise}
\end{eqnarray}
We are setting $k_{\text{B}}=1$ and the kinetic coefficient
$\Gamma=1$; note that for the motion in a viscous medium we have
according to Stokes theorem $\Gamma=1/6\pi\eta R$, where $R$ is the
radius of the particle and $\eta$ the viscosity.

The kinetics of a particle moving in a general potential is
complex. On the one hand, the particle can be trapped in local
minima on a time scale given by the inverse spring constant in a
local harmonic approximation; on the other hand, the particle can
also make a Kramers transition across potential barriers
separating local minima. These transitions are typically dominated
by the Arrhenius factor $\exp(-\Delta U/T)$, where $\Delta U$ is
the potential barrier. However, in the long time limit, i.e., at
times larger than the largest relaxation time, the particle
samples the full potential profile and the stationary distribution
is given by the Boltzmann expression
\begin{eqnarray}
P_0(x)=\frac{e^{-\beta U(x)}}{Z(\beta)}, \label{stat}
\end{eqnarray}
where $Z(\beta)$ is the partition function
\begin{eqnarray}
Z(\beta)=\int dxe^{-\beta U(x)}, \label{par}
\end{eqnarray}
and $\beta=1/T$ (the inverse temperature).

At low temperature, $\beta\rightarrow\infty$, a steepest descent
argument applied to the partition function (\ref{par}) implies
that only the local minima in $U$ contribute. The particle is
trapped in the local minima and only rarely makes  Kramers
transitions to neighboring wells. Expanding the potential to
quadratic order about the well at position $x_i$ with gap $U_i$
and second derivative $k_i$ (spring constant) we obtain the local
contribution
\begin{eqnarray}
U(x)\sim U_i+\frac{1}{2}k_i(x-x_i)^2, \label{harmonic}
\end{eqnarray}
and for the partition function, performing the Gaussian integral,
$\int dx\exp(-\alpha x^2)=\sqrt{\pi/\alpha}$,
\begin{eqnarray}
Z(\beta)\sim\sum_i\left(\frac{2\pi}{\beta k_i}\right)^{1/2}
e^{-\beta U_i}. \label{parhar}
\end{eqnarray}

In the present letter we wish to focus on the distribution of the
heat exchange $Q(t)$ with the reservoir in the long time - low
temperature limit. For the Brownian motion of a particle in a
potential the heat $Q(t)$ sampled up to time $t$ is a fluctuating
quantity characterized by a time dependent probability distribution
$P(Q,t)$. In the long time limit the distribution $P(Q,t)$
approaches a stationary distribution $P_0(Q)$ which we proceed to
analyze.

The heat delivered by the heat bath in the time span $t$ is in
general given by the expression \cite{Sekimoto97,AI06,AI07}
\begin{eqnarray}
Q(t)=\int_{x_0}^{x(t)} dx
\frac{dU}{dx}=\int_0^tdt'\left(\frac{dU}{dx}\right)_{x(t')}\frac{dx(t')}{dt'},
\label{heat}
\end{eqnarray}
where the first integral is interpreted according to the
Stratonovich integration scheme and where we simply sample the
energy change associated with the fluctuating position of the
particle; $x_0$ is the initial position and $x(t)$ the position at
time $t$. The expression (\ref{heat}) can, of course, in principle
be evaluated by the insertion of (\ref{lan}). However, this more
cumbersome procedure can be circumvented by noting that simple
quadrature yields
\begin{eqnarray}
Q(t)=U(x(t))-U(x_0).\label{intheat}
\end{eqnarray}
Note that this result only holds for a static potential. In the case
of a prescribed dynamic potential doing work on the particle
according to a given protocol, the expression (\ref{intheat}) does
not hold and one is faced with the more intractable problem of
handling (\ref{heat}).

Introducing the characteristic function the heat distribution is
given by
\begin{eqnarray}
P(Q,t)=\int\frac{dp}{2\pi}e^{ipQ}\langle e^{ -ipQ(t)}\rangle,
\label{dis}
\end{eqnarray}
where the average $\langle\cdots\rangle$ in the characteristic
function is both over the initial position $x_0$ and the final
position $x(t)$. At $t=0$ the heat distribution $P(Q,0)=\delta(Q)$.
After a transient period of order the inverse spring constants and
Kramers rates the heat distribution function becomes stationary. The
particle is at all times in thermal equilibrium characterized by the
Boltzmann distribution (\ref{stat}). Accordingly, averaging at long
times over $x_0$ and $x(\infty)$ using (\ref{stat}) we obtain for
the characteristic function
\begin{eqnarray}
\langle e^{-ipQ(\infty)}\rangle= \int dx_0dx P_0(x_0)P_0(x)
e^{-ipU(x)+ipU(x_0)}.~~~~~~ \label{charq}
\end{eqnarray}
Finally, extending the partition function (\ref{par}) to complex
inverse temperature we obtain in a compact manner a general long
time expression for the characteristic function
\begin{eqnarray}
\langle e^{-ipQ(\infty)}\rangle=\frac{|Z(\beta+ip)|^2}{Z(\beta)^2};
\label{chargen}
\end{eqnarray}
note that $\langle e^{-ipQ(\infty)}\rangle_{p=0}=1$ ensuring the
normalization condition $\int dQ P(Q,\infty)=1$.

At low temperature the partition function predominantly samples the
local minima in $U$ and we obtain inserting the asymptotic
expression (\ref{parhar}) generalized to complex inverse temperature
the low temperature - long time expression for the characteristic
function
\begin{eqnarray}
\langle e^{-ipQ(\infty)}\rangle=
\left[\frac{\beta^2}{\beta^2+p^2}\right]^{1/2}
\frac{\sum_{ij}(k_ik_j)^{-1/2}\exp(-\beta(U_i+U_j))\exp(-ip(U_i-U_j))}
{\sum_{nm}(k_nk_m)^{-1/2}\exp(-\beta(U_n+U_m))}. \label{charhar}
\end{eqnarray}
First, we note again that $\langle e^{-ipQ(\infty)}\rangle_{p=0}=1$
yielding the normalization of $P(Q,\infty)$, moreover, the phase
factors $\exp(-ip(U_i-U_j))$ can according to (\ref{dis}) be
absorbed in a shift of $Q$. The interesting aspect resides in the
prefactor $(\beta^2+p^2)^{-1/2}$ which has branch points in the
complex $p$ plane at $p=\pm i\beta$.

By inspection of (\ref{dis}) and (\ref{charhar}) we note that for
small $Q$ relative to $U_i-U_j$ the integral in (\ref{dis}) is
logarithmically divergent for large $p$ since
$(\beta^2+p^2)^{-1/2}\sim p^{-1}$, implying a logarithmically
divergent contribution to $P(Q,\infty)\equiv P_0(Q)$, i.e.,
$P_0(Q)\sim -\log|Q-(U_i-U_j)|$. For large $|Q|$, i.e., the tails of
the distribution $P_0(Q)$, we sample the small $p$ region in
(\ref{dis}) and (\ref{charhar}) and closing the contour in the upper
half plane (lower half plane) for $Q>0$ ($Q<0$) picking up the
branch point contribution $p=i\beta$ ($p=-i\beta$) we obtain the
dominant exponential tails $P_0(Q)\sim \exp(-\beta|Q|)$.

However, using the well-known identity \cite{Gradshteyn65}
\begin{eqnarray}
\int_0^\infty dx\frac{\cos ax}{(b^2+x^2)^{1/2}}=K_0(ab),
\label{bessel}
\end{eqnarray}
where $K_0(x)$ is a Bessel function of the second kind, it is easy
to derive an explicit expression for the heat distribution
function. We have
\begin{eqnarray}
P_0(Q)=\frac{\beta}{\pi}
\frac{\sum_{ij}(k_ik_j)^{-1/2}e^{-\beta(U_i+U_j)}
K_0(\beta|Q-(U_i-U_j)|)}{\sum_{nm}(k_nk_m)^{-1/2}e^{-\beta(U_n+U_m)}}.
\label{gen}
\end{eqnarray}
Using the identity $\int dx K_0(x)=\pi$ following from
(\ref{bessel}) we confirm the normalization condition $\int
dQP_0(Q)=1$. For small argument $K_0(x)\sim -\log(x)$ and we obtain
for $Q\sim U_i-U_j$
\begin{eqnarray}
P_0(Q)\approx
-\frac{\beta}{\pi}\frac{(k_ik_j)^{-1/2}e^{-\beta(U_i+U_j)}}
{\sum_{nm}(k_nk_m)^{-1/2}e^{-\beta(U_n+U_m)}}\log|Q-(U_i-U_j)|,
\label{gendiv}
\end{eqnarray}
showing that $P_0(Q)$ exhibits a multi band structure of log
divergent peaks at $Q=U_i-U_j$ in agreement with our qualitative
discussion. For large argument $K_0(x)\sim(\pi/2x)^{1/2}\exp(-x)$
and we obtain the exponential tails $P_0(Q)\sim
Q^{-1/2}\exp(-\beta|Q|)$, including the prefactor $Q^{-1/2}$.

\begin{figure}
\includegraphics[width=1.0\hsize]{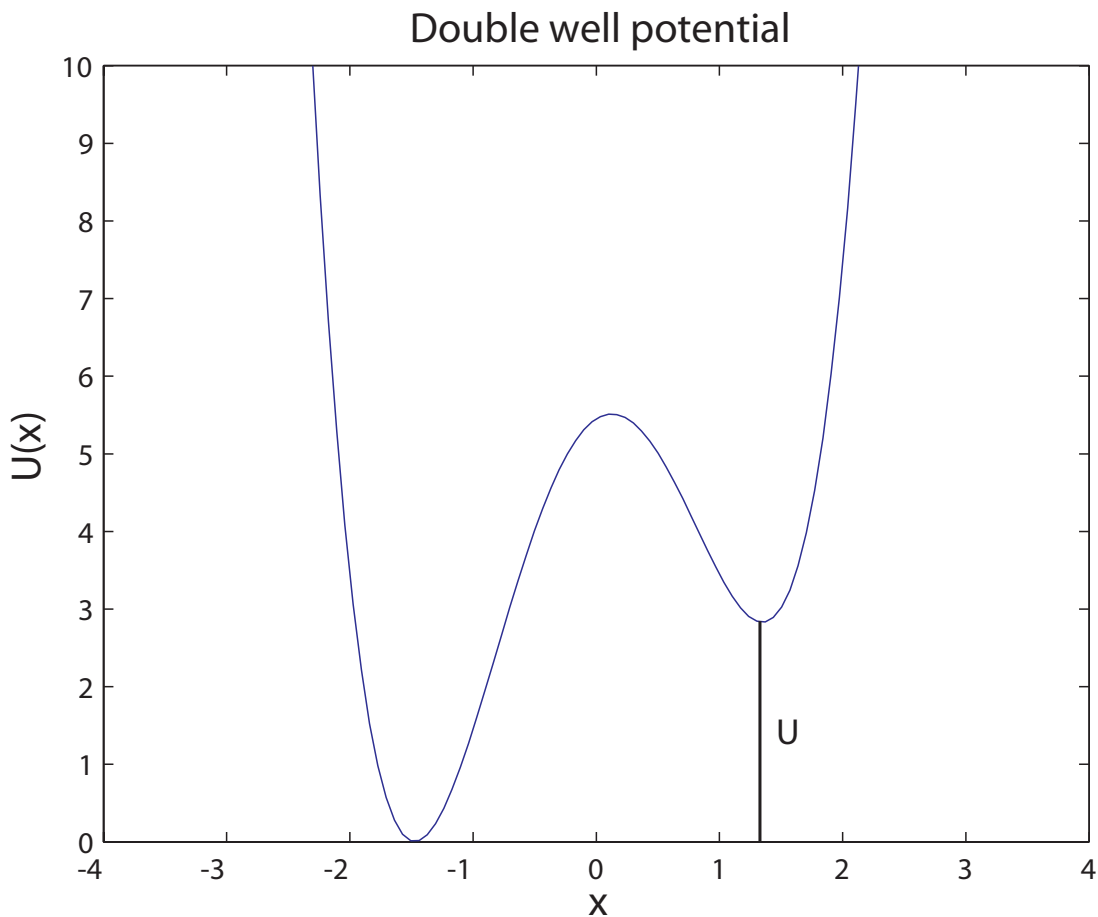}
\caption{Double well potential with gap $U$ (arbitrary units).}
\label{fig1}
\end{figure}

\begin{figure}
\includegraphics[width=1.0\hsize]{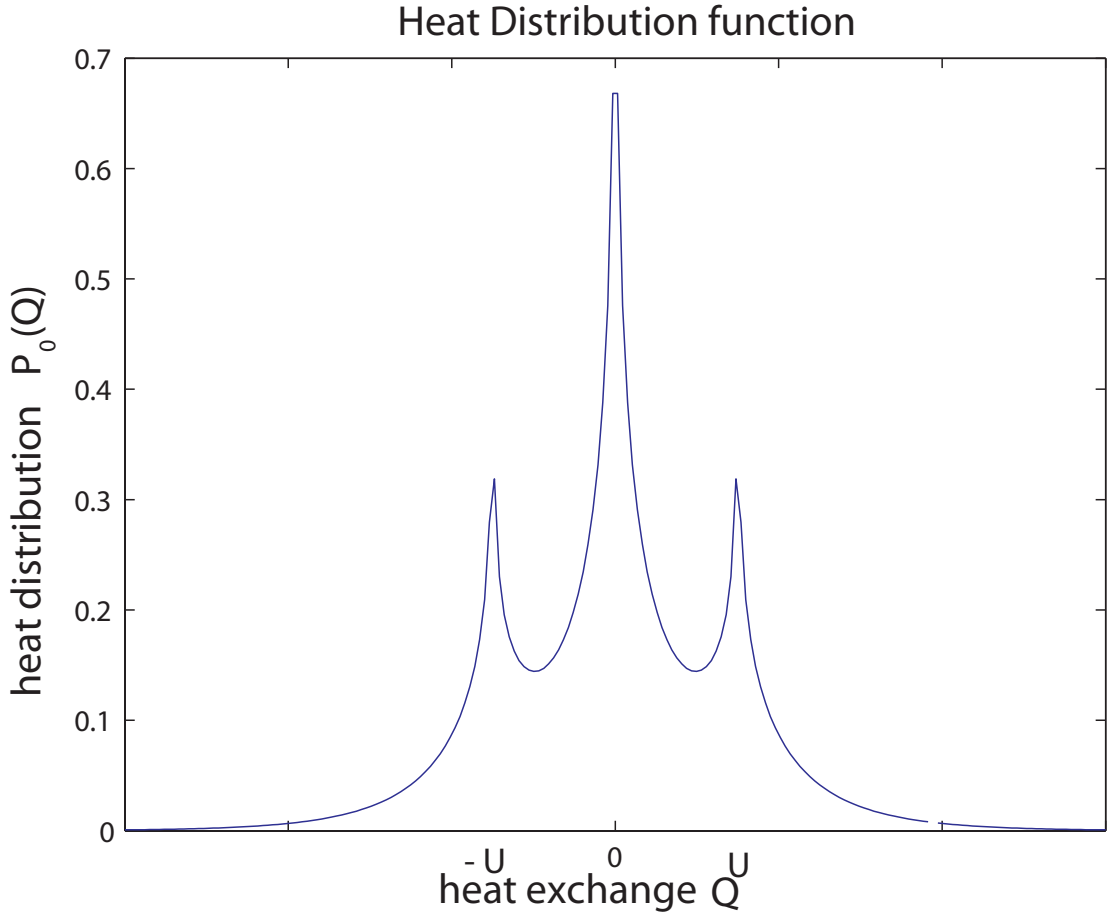}
\caption{Heat distribution function pertaining to a double well
potential with gap $U$. The spectrum shows a central peak at $Q=0$
and two side bands at $Q=\pm U$ (arbitrary units).} \label{fig2}
\end{figure}

In the case of the double well potential depicted in Fig.~\ref{fig1}
with two minima, the second minimum with gap $U$, we obtain from
(\ref{charhar}) the characteristic function
\begin{eqnarray}
\langle e^{-ipQ(\infty)}\rangle=
\left[\frac{\beta^2}{\beta^2+p^2}\right]^{1/2}
\left[1+2\left(\frac{k_1}{k_2}\right)^{1/2}e^{-\beta U}(\cos pU -1)
\right], \label{chardw}
\end{eqnarray}
yielding the heat distribution function
\begin{eqnarray}
P_0(Q)=&&\frac{\beta}{\pi}\left[1-2\left(\frac{k_1}{k_2}\right)^{1/2}
e^{-\beta U}\right]K_0(\beta|Q|) \nonumber
\\
+&&\frac{\beta}{\pi}\left(\frac{k_1}{k_2}\right)^{1/2}e^{-\beta U}
(K_0(\beta|Q-U|) +K_0(\beta|Q+U|)). \label{disdw}
\end{eqnarray}
In Fig.~\ref{fig2} we have shown the heat distribution function
$P_0(Q)$ as a function of heat transfer $Q$ for the parameter
values $k_1=k_2=1$, $\beta = 1$, and $U=1.5$.

The interpretation of the log divergent multi band structure in the
heat distribution function in the low temperature-long time limit is
easy. From the work of Imparato el al. \cite{Imparato07}, see also
the work of van Zon et al. \cite{vanZon03}, it is well-known that
the heat distribution function for a static harmonic potential is
given by the Bessel function $K_0(\beta|Q|)$ exhibiting a log
divergence at zero heat transfer $Q\sim 0$; for large $Q$ the
distribution falls off according to the Boltzmann factor
$\exp(-\beta|Q|)$. For a general potential possessing several minima
these features seem to persist. Each minima in the potential acts
like a local reservoir where the particle at low temperature can be
trapped for a long time before making a Kramers transition to
another well. Since the transfer between the potential well with gap
$U_i$ to the well with gap $U_j$ involves the energy difference
$U_i-U_j$ the divergent band appear at heat transfer $Q=U_i-U_j$. It
also follows from (\ref{gendiv}) that the contributions are weighted
with the corresponding Boltzmann factors.

In the case of the double well potential depicted in Fig.~\ref{fig1}
with a well with zero gap and a well with gap $U$ the discussion is
particularly transparent. At low temperature the particle is for
most of the time trapped in the zero gap well yielding the log
divergent behavior for zero heat transfer. Occasionally the particle
makes a Kramers transition to the well with gap $U$ and becomes
trapped yielding log divergent peaks at $Q\sim\pm U$. The side bands
originating from the well with the gap are down by the Boltmann
factor $\exp(-\beta U)$. The sum rule (normalization) $\int dQ
P_0(q)=1$ implies that the total integrated strength is constant. We
also note that for the case of a vanishing gap $U=0$, i.e., for two
gapless wells, we recover the result
$P_0(Q)=(\beta/\pi)K_0(\beta|Q|)$, independent of the spring
constant \cite{Imparato07}.

\begin{figure}
\includegraphics[width=1.0\hsize]{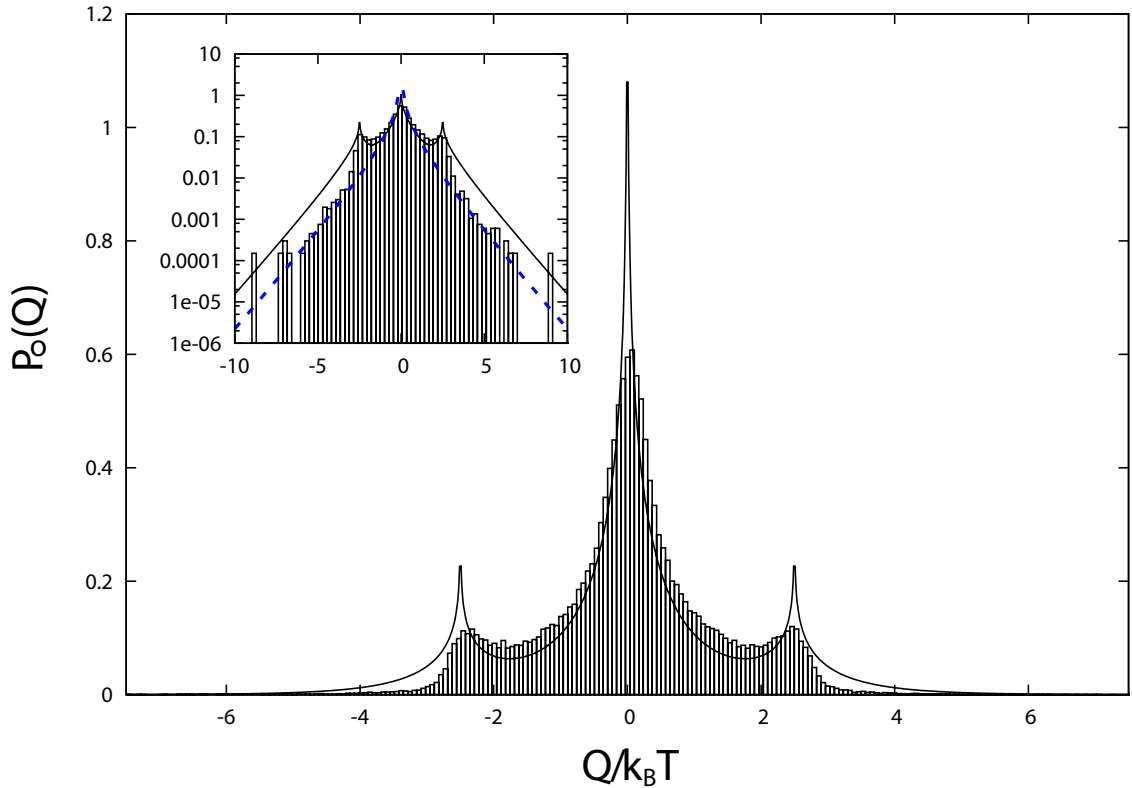}
\caption{Analytical solution eq.~(\ref{disdw}) (full line), and heat
distribution (histogram) as obtained by numerical solution of the
Langevin equation ($10^5$ independent trajectories, with $t_0=0$,
$t=10$), for the potential in (\ref{pol}) with $d=15$, $U=2.5$,
$\beta=1$. Inset: Heat distribution in log-linear scale. Dashed
lines: $|Q|^{-3/4}\exp(-\beta |Q|)$.} \label{fig3}
\end{figure}

In order to model a double well potential in more detail we have
used the fourth order polynomial
\begin{eqnarray}
U(x)=dx^4+4Ux^3+\left(\frac{9U^2}{2d}-\frac{d}{2}\right)x^2,
\label{pol}
\end{eqnarray}
which has a gap $U$, minima located at $x=-(3U/2d)\pm 1/2$, and
spring constants $k=2(d\pm 3U)$. Choosing $\beta=1$, $U=2.5$ and
$d=15$ we have in Fig.~\ref{fig3} depicted the analytical expression
(\ref{disdw}) together with a numerical solution of the Langevin
equation (($10^5$ independent trajectories, with $t_0=0$, $t=10$).

Finally, we derive a general expression for the large $|Q|$ behavior
of the heat distribution function $P_0(Q)$. For large $|Q|$,
corresponding to large heat transfer to the heat bath, we sample the
wings of the potential. Considering a general potential $U$ behaving
like $U\approx Ax^n$, for large $|x|$, $n$ even, we obtain for the
heat distribution function
\begin{eqnarray}
P_0(Q)\propto \int dx_1dx_2e^{-\beta(U_1+U_2)}\delta(|Q|-U_1+U_2)
\label{heatdis}
\end{eqnarray}
where $U_1=U(x_1)$ and $U_2=U(x_2)$. For large $|Q|$ inserting
$U\approx Ax^n$ and introducing polar coordinates $x_1=r\cos\phi$,
$x_2=r\sin\phi$ we have, using the delta function to eliminate $r$,
\begin{eqnarray}
P_0(Q)\propto \int rdrd\phi e^{-\beta |Q|F(\phi)}
\delta(|Q|-Ar^n(\cos^n\phi-\sin^n\phi)), \label{heatdis2}
\end{eqnarray}
where $F(\phi)=(1+\tan^n\phi)/(1-\tan^n\phi)$. For large $Q$ the
integral is dominated by the minima of $F$ for $\phi=0$ and
$\phi=\pi$. Expanding about the minima to second order, $F\approx
1-2(\delta\phi)^n$, performing the Gaussian integrals and the
r-integration over the delta function in (\ref{heatdis2}) we obtain
the distribution function for large $|Q|$
\begin{eqnarray}
P_0(Q)\propto |Q|^{1/n-1}e^{-\beta |Q|}. \label{heatdis3}
\end{eqnarray}
This is a general result. Here $e^{-\beta|Q|}$ is a Boltzmann
factor associated with the heat transfer $Q$ whereas the prefactor
$|Q|^{1/n-1}$ is a "density of states" contribution. For $n=2$ we
obtain the previous Bessel result, $P_0(Q)\approx
|Q|^{-1/2}\exp(-\beta|Q|)$, pertaining to the harmonic
approximation, for $n=4$, corresponding to the double well model
potential (\ref{pol}), we have $P_0(Q)\approx
|Q|^{-1/2}\exp(-\beta |Q|)$. In the inset in Fig.~\ref{fig3} we
have depicted the heat distribution in a log-linear scale. The
dashed line corresponds to $P_0(Q)\propto
|Q|^{-3/4}\exp(-\beta|Q|)$.

In this letter we have generalized the result for the heat
distribution function for a harmonic potential to the case of a
general static potential with several minima in the long time - low
temperature limit. Our analysis shows that the gap structure of the
potential wells give rise to a multi band structure of log divergent
peaks in the heat distribution function. We have, moreover, derived
a general result for the tails of the heat distribution function.
\acknowledgments

The work of H. Fogedby has been supported by the Danish Natural
Science Research Council under grant no. 95093801. The work of A.
Imparato has been supported by a grant from the Department of
Physics and Astronomy. We would like to thank A. Svane for useful
discussions.

\begin{thebibliography}{30}
\expandafter\ifx\csname
natexlab\endcsname\relax\def\natexlab#1{#1}\fi
\expandafter\ifx\csname bibnamefont\endcsname\relax
  \def\bibnamefont#1{#1}\fi
\expandafter\ifx\csname bibfnamefont\endcsname\relax
  \def\bibfnamefont#1{#1}\fi
\expandafter\ifx\csname citenamefont\endcsname\relax
  \def\citenamefont#1{#1}\fi
\expandafter\ifx\csname url\endcsname\relax
  \def\url#1{\texttt{#1}}\fi
\expandafter\ifx\csname urlprefix\endcsname\relax\def\urlprefix{URL
}\fi \providecommand{\bibinfo}[2]{#2}
\providecommand{\eprint}[2][]{\url{#2}}

\bibitem[{\citenamefont{Trepagnier et~al.}(2004)\citenamefont{Trepagnier,
  Jarzynski, Ritort, Crooks, Bustamante, and Liphardt}}]{Felix04}
\bibinfo{author}{\bibfnamefont{E.}~\bibnamefont{Trepagnier}},
  \bibinfo{author}{\bibfnamefont{C.}~\bibnamefont{Jarzynski}},
  \bibinfo{author}{\bibfnamefont{F.}~\bibnamefont{Ritort}},
  \bibinfo{author}{\bibfnamefont{G.}~\bibnamefont{Crooks}},
  \bibinfo{author}{\bibfnamefont{C.}~\bibnamefont{Bustamante}},
  \bibnamefont{and} \bibinfo{author}{\bibfnamefont{J.}~\bibnamefont{Liphardt}},
  \bibinfo{journal}{Proc. Natl. Acad. Sci. USA} \textbf{\bibinfo{volume}{101}},
  \bibinfo{pages}{15038} (\bibinfo{year}{2004}).

\bibitem[{\citenamefont{Collin et~al.}(2005)\citenamefont{Collin, Ritort,
  Jarzynski, Smith, Jr, and Bustamante}}]{Felix05}
\bibinfo{author}{\bibfnamefont{D.}~\bibnamefont{Collin}},
  \bibinfo{author}{\bibfnamefont{F.}~\bibnamefont{Ritort}},
  \bibinfo{author}{\bibfnamefont{C.}~\bibnamefont{Jarzynski}},
  \bibinfo{author}{\bibfnamefont{S.~B.} \bibnamefont{Smith}},
  \bibinfo{author}{\bibfnamefont{I.~T.} \bibnamefont{Jr}}, \bibnamefont{and}
  \bibinfo{author}{\bibfnamefont{C.}~\bibnamefont{Bustamante}},
  \bibinfo{journal}{Nature} \textbf{\bibinfo{volume}{437}},
  \bibinfo{pages}{231} (\bibinfo{year}{2005}).

\bibitem[{\citenamefont{Tietz et~al.}(2006)\citenamefont{Tietz, Schuler, Speck,
  Seifert, and Wrachtrup}}]{Seifert06a}
\bibinfo{author}{\bibfnamefont{C.}~\bibnamefont{Tietz}},
  \bibinfo{author}{\bibfnamefont{S.}~\bibnamefont{Schuler}},
  \bibinfo{author}{\bibfnamefont{T.}~\bibnamefont{Speck}},
  \bibinfo{author}{\bibfnamefont{U.}~\bibnamefont{Seifert}}, \bibnamefont{and}
  \bibinfo{author}{\bibfnamefont{J.}~\bibnamefont{Wrachtrup}},
  \bibinfo{journal}{Phys. Rev. Lett.} \textbf{\bibinfo{volume}{97}},
  \bibinfo{pages}{050602} (\bibinfo{year}{2006}).

\bibitem[{\citenamefont{Blickle et~al.}(2006)\citenamefont{Blickle, Speck,
  Helden, U.Seifert, and Bechinger}}]{Seifert06b}
\bibinfo{author}{\bibfnamefont{V.}~\bibnamefont{Blickle}},
  \bibinfo{author}{\bibfnamefont{T.}~\bibnamefont{Speck}},
  \bibinfo{author}{\bibfnamefont{L.}~\bibnamefont{Helden}},
  \bibinfo{author}{\bibnamefont{U.Seifert}}, \bibnamefont{and}
  \bibinfo{author}{\bibfnamefont{C.}~\bibnamefont{Bechinger}},
  \bibinfo{journal}{Phys. Rev. Lett.} \textbf{\bibinfo{volume}{96}},
  \bibinfo{pages}{070603} (\bibinfo{year}{2006}).

\bibitem[{\citenamefont{Wang et~al.}(2002)\citenamefont{Wang, Sevick, Mittag,
  Searles, and Evans}}]{Wang02}
\bibinfo{author}{\bibfnamefont{G.}~\bibnamefont{Wang}},
  \bibinfo{author}{\bibfnamefont{E.}~\bibnamefont{Sevick}},
  \bibinfo{author}{\bibfnamefont{E.}~\bibnamefont{Mittag}},
  \bibinfo{author}{\bibfnamefont{D.~J.} \bibnamefont{Searles}},
  \bibnamefont{and} \bibinfo{author}{\bibfnamefont{D.~J.} \bibnamefont{Evans}},
  \bibinfo{journal}{Phys. Rev. Lett.} \textbf{\bibinfo{volume}{89}},
  \bibinfo{pages}{050601} (\bibinfo{year}{2002}).

\bibitem[{\citenamefont{Imparato
  et~al.}(2007{\natexlab{a}})\citenamefont{Imparato, Peliti, Pesce, Rusciano,
  and Sasso}}]{AI07}
\bibinfo{author}{\bibfnamefont{A.}~\bibnamefont{Imparato}},
  \bibinfo{author}{\bibfnamefont{L.}~\bibnamefont{Peliti}},
  \bibinfo{author}{\bibfnamefont{G.}~\bibnamefont{Pesce}},
  \bibinfo{author}{\bibfnamefont{G.}~\bibnamefont{Rusciano}}, \bibnamefont{and}
  \bibinfo{author}{\bibfnamefont{A.}~\bibnamefont{Sasso}},
  \bibinfo{journal}{Phys. Rev. E} \textbf{\bibinfo{volume}{76}},
  \bibinfo{pages}{050101R} (\bibinfo{year}{2007}{\natexlab{a}}).

\bibitem[{\citenamefont{Douarche et~al.}(2006)\citenamefont{Douarche, Joubaud,
  Garnier, Petrosyan, and Ciliberto}}]{Ciliberto06}
\bibinfo{author}{\bibfnamefont{F.}~\bibnamefont{Douarche}},
  \bibinfo{author}{\bibfnamefont{S.}~\bibnamefont{Joubaud}},
  \bibinfo{author}{\bibfnamefont{N.~B.} \bibnamefont{Garnier}},
  \bibinfo{author}{\bibfnamefont{A.}~\bibnamefont{Petrosyan}},
  \bibnamefont{and}
  \bibinfo{author}{\bibfnamefont{S.}~\bibnamefont{Ciliberto}},
  \bibinfo{journal}{Phys. Rev. Lett.} \textbf{\bibinfo{volume}{97}},
  \bibinfo{pages}{140603} (\bibinfo{year}{2006}).

\bibitem[{\citenamefont{Garnier and Ciliberto}(2007)}]{Ciliberto07}
\bibinfo{author}{\bibfnamefont{N.}~\bibnamefont{Garnier}} \bibnamefont{and}
  \bibinfo{author}{\bibfnamefont{S.}~\bibnamefont{Ciliberto}},
  \bibinfo{journal}{Phys. Rev. E} \textbf{\bibinfo{volume}{71}},
  \bibinfo{pages}{060101(R)} (\bibinfo{year}{2007}).

\bibitem[{\citenamefont{Imparato et~al.}(2008)\citenamefont{Imparato, Jop,
  Petrosyan, and Ciliberto}}]{Ciliberto08}
\bibinfo{author}{\bibfnamefont{A.}~\bibnamefont{Imparato}},
  \bibinfo{author}{\bibfnamefont{P.}~\bibnamefont{Jop}},
  \bibinfo{author}{\bibfnamefont{A.}~\bibnamefont{Petrosyan}},
  \bibnamefont{and}
  \bibinfo{author}{\bibfnamefont{S.}~\bibnamefont{Ciliberto}},
  \bibinfo{journal}{J. Stat. Mech} p. \bibinfo{pages}{P10017}
  (\bibinfo{year}{2008}).

\bibitem[{\citenamefont{Jarzynski}(1997)}]{Jarzynski97}
\bibinfo{author}{\bibfnamefont{C.}~\bibnamefont{Jarzynski}},
  \bibinfo{journal}{Phys. Rev. Lett.} \textbf{\bibinfo{volume}{78}},
  \bibinfo{pages}{2690} (\bibinfo{year}{1997}).

\bibitem[{\citenamefont{Kurchan}(1998)}]{Kurchan98}
\bibinfo{author}{\bibfnamefont{J.}~\bibnamefont{Kurchan}}, \bibinfo{journal}{J.
  Phys. A} \textbf{\bibinfo{volume}{31}}, \bibinfo{pages}{3719}
  (\bibinfo{year}{1998}).

\bibitem[{\citenamefont{Gallavotti}(1996)}]{Gallavotti96}
\bibinfo{author}{\bibfnamefont{G.}~\bibnamefont{Gallavotti}},
  \bibinfo{journal}{Phys. Rev. Lett.} \textbf{\bibinfo{volume}{77}},
  \bibinfo{pages}{4334} (\bibinfo{year}{1996}).

\bibitem[{\citenamefont{Crooks}(1999)}]{Crooks99}
\bibinfo{author}{\bibfnamefont{G.~E.} \bibnamefont{Crooks}},
  \bibinfo{journal}{Phys. Rev. E} \textbf{\bibinfo{volume}{60}},
  \bibinfo{pages}{2721} (\bibinfo{year}{1999}).

\bibitem[{\citenamefont{Crooks}(2000)}]{Crooks00}
\bibinfo{author}{\bibfnamefont{G.~E.} \bibnamefont{Crooks}},
  \bibinfo{journal}{Phys. Rev. E} \textbf{\bibinfo{volume}{61}},
  \bibinfo{pages}{2361} (\bibinfo{year}{2000}).

\bibitem[{\citenamefont{Seifert}(2005{\natexlab{a}})}]{Seifert05a}
\bibinfo{author}{\bibfnamefont{U.}~\bibnamefont{Seifert}},
  \bibinfo{journal}{Phys. Rev. Lett.} \textbf{\bibinfo{volume}{95}},
  \bibinfo{pages}{040602} (\bibinfo{year}{2005}{\natexlab{a}}).

\bibitem[{\citenamefont{Seifert}(2005{\natexlab{b}})}]{Seifert05b}
\bibinfo{author}{\bibfnamefont{U.}~\bibnamefont{Seifert}},
  \bibinfo{journal}{Europhys. Lett} \textbf{\bibinfo{volume}{70}},
  \bibinfo{pages}{36} (\bibinfo{year}{2005}{\natexlab{b}}).

\bibitem[{\citenamefont{Evans et~al.}(1993)\citenamefont{Evans, Cohen, and
  Morriss}}]{Evans93}
\bibinfo{author}{\bibfnamefont{D.~J.} \bibnamefont{Evans}},
  \bibinfo{author}{\bibfnamefont{E.~G.~D.} \bibnamefont{Cohen}},
  \bibnamefont{and} \bibinfo{author}{\bibfnamefont{G.~P.}
  \bibnamefont{Morriss}}, \bibinfo{journal}{Phys. Rev. Lett.}
  \textbf{\bibinfo{volume}{71}}, \bibinfo{pages}{2401} (\bibinfo{year}{1993}).

\bibitem[{\citenamefont{Evans and Searles}(1994)}]{Evans94}
\bibinfo{author}{\bibfnamefont{D.~J.} \bibnamefont{Evans}} \bibnamefont{and}
  \bibinfo{author}{\bibfnamefont{D.~J.} \bibnamefont{Searles}},
  \bibinfo{journal}{Phys. Rev. E} \textbf{\bibinfo{volume}{50}},
  \bibinfo{pages}{1645} (\bibinfo{year}{1994}).

\bibitem[{\citenamefont{Gallavotti and Cohen}(1995)}]{GC95}
\bibinfo{author}{\bibfnamefont{G.}~\bibnamefont{Gallavotti}} \bibnamefont{and}
  \bibinfo{author}{\bibnamefont{Cohen}}, \bibinfo{journal}{J. Stat. Phys.}
  \textbf{\bibinfo{volume}{80}}, \bibinfo{pages}{931} (\bibinfo{year}{1995}).

\bibitem[{\citenamefont{Lebowitz and Spohn}(1999)}]{LS99}
\bibinfo{author}{\bibfnamefont{J.~L.} \bibnamefont{Lebowitz}} \bibnamefont{and}
  \bibinfo{author}{\bibfnamefont{H.}~\bibnamefont{Spohn}}, \bibinfo{journal}{J.
  Stat. Phys.} \textbf{\bibinfo{volume}{95}}, \bibinfo{pages}{333}
  (\bibinfo{year}{1999}).

\bibitem[{\citenamefont{Gaspard}(2004)}]{Gaspard04}
\bibinfo{author}{\bibfnamefont{P.}~\bibnamefont{Gaspard}}, \bibinfo{journal}{J.
  Stat. Phys.} \textbf{\bibinfo{volume}{117}}, \bibinfo{pages}{599}
  (\bibinfo{year}{2004}).

\bibitem[{\citenamefont{Imparato and Peliti}(2006)}]{AI06}
\bibinfo{author}{\bibfnamefont{A.}~\bibnamefont{Imparato}} \bibnamefont{and}
  \bibinfo{author}{\bibfnamefont{L.}~\bibnamefont{Peliti}},
  \bibinfo{journal}{Phys. Rev. E} \textbf{\bibinfo{volume}{74}},
  \bibinfo{pages}{026106} (\bibinfo{year}{2006}).

\bibitem[{\citenamefont{van Zon and Cohen}(2003{\natexlab{a}})}]{vanZon03}
\bibinfo{author}{\bibfnamefont{R.}~\bibnamefont{van Zon}} \bibnamefont{and}
  \bibinfo{author}{\bibfnamefont{E.~G.~D.} \bibnamefont{Cohen}},
  \bibinfo{journal}{Phys. Rev. Lett.} \textbf{\bibinfo{volume}{91}},
  \bibinfo{pages}{110601} (\bibinfo{year}{2003}{\natexlab{a}}).

\bibitem[{\citenamefont{van Zon et~al.}(2004)\citenamefont{van Zon, Ciliberto,
  and Cohen}}]{vanZon04}
\bibinfo{author}{\bibfnamefont{R.}~\bibnamefont{van Zon}},
  \bibinfo{author}{\bibfnamefont{S.}~\bibnamefont{Ciliberto}},
  \bibnamefont{and} \bibinfo{author}{\bibfnamefont{E.~G.~D.}
  \bibnamefont{Cohen}}, \bibinfo{journal}{Phys. Rev. Lett.}
  \textbf{\bibinfo{volume}{92}}, \bibinfo{pages}{130601}
  (\bibinfo{year}{2004}).

\bibitem[{\citenamefont{van Zon and Cohen}(2003{\natexlab{b}})}]{vanZon03a}
\bibinfo{author}{\bibfnamefont{R.}~\bibnamefont{van Zon}} \bibnamefont{and}
  \bibinfo{author}{\bibfnamefont{E.~G.~D.} \bibnamefont{Cohen}},
  \bibinfo{journal}{Phys. Rev.} \textbf{\bibinfo{volume}{67}},
  \bibinfo{pages}{046102} (\bibinfo{year}{2003}{\natexlab{b}}).

\bibitem[{\citenamefont{van Zon and Cohen}(2004)}]{vanZon04a}
\bibinfo{author}{\bibfnamefont{R.}~\bibnamefont{van Zon}} \bibnamefont{and}
  \bibinfo{author}{\bibfnamefont{E.~G.~D.} \bibnamefont{Cohen}},
  \bibinfo{journal}{Phys. Rev. E} \textbf{\bibinfo{volume}{69}},
  \bibinfo{pages}{056121} (\bibinfo{year}{2004}).

\bibitem[{\citenamefont{Speck and Seifert}(2005)}]{Seifert05c}
\bibinfo{author}{\bibfnamefont{T.}~\bibnamefont{Speck}} \bibnamefont{and}
  \bibinfo{author}{\bibfnamefont{U.}~\bibnamefont{Seifert}},
  \bibinfo{journal}{Eur. Phys. J. B} \textbf{\bibinfo{volume}{43}},
  \bibinfo{pages}{521} (\bibinfo{year}{2005}).

\bibitem[{Sek(1997)}]{Sekimoto97}
\bibinfo{journal}{J. Phys. Soc. Jpn} \textbf{\bibinfo{volume}{66}},
  \bibinfo{pages}{1234} (\bibinfo{year}{1997}).

\bibitem[{\citenamefont{Gradshteyn and Ryzhik}(1965)}]{Gradshteyn65}
\bibinfo{author}{\bibfnamefont{I.~S.} \bibnamefont{Gradshteyn}}
  \bibnamefont{and} \bibinfo{author}{\bibfnamefont{I.~M.}
  \bibnamefont{Ryzhik}}, \emph{\bibinfo{title}{Table of Integrals. Series, and
  Products}} (\bibinfo{publisher}{Academic Press}, \bibinfo{address}{New York},
  \bibinfo{year}{1965}).

\bibitem[{\citenamefont{Imparato
  et~al.}(2007{\natexlab{b}})\citenamefont{Imparato, Peliti, Pesce, Rusciano,
  and Sasso}}]{Imparato07}
\bibinfo{author}{\bibfnamefont{A.}~\bibnamefont{Imparato}},
  \bibinfo{author}{\bibfnamefont{L.}~\bibnamefont{Peliti}},
  \bibinfo{author}{\bibfnamefont{G.}~\bibnamefont{Pesce}},
  \bibinfo{author}{\bibfnamefont{G.}~\bibnamefont{Rusciano}}, \bibnamefont{and}
  \bibinfo{author}{\bibfnamefont{A.}~\bibnamefont{Sasso}},
  \bibinfo{journal}{Phys. Rev. E} \textbf{\bibinfo{volume}{76}},
  \bibinfo{pages}{050101} (\bibinfo{year}{2007}{\natexlab{b}}).

\end{thebibliography}

\end{document}